%% file: main.tex
\documentclass[runningheads]{llncs}

\usepackage[T1]{fontenc}
\usepackage[utf8]{inputenc}
\usepackage{microtype}
\usepackage{inconsolata}
\usepackage{graphicx}

\usepackage{amsmath}
\usepackage{amssymb}
\usepackage{bm}
\usepackage{booktabs}
\usepackage{multirow}
\usepackage{makecell}
\usepackage{xcolor}
\usepackage{xspace}
\usepackage{algorithm}
\usepackage{algpseudocode}
\usepackage{enumitem}
\usepackage{subcaption}
\usepackage{tabularx}
\usepackage{hyperref}
\usepackage{float}

\begin{document}

\title{
The Impact of Backbone Evolution on LLM-Based Relevance Assessments}
\titlerunning{The impact of backbone evolution on LLM-based relevance assessments}

 \author{ Chuting Yu\orcidID{0009-0008-5150-3843} \and \\
Guido Zuccon\orcidID{0000-0003-0271-5563} \and \\
Teerapong Leelanupab\orcidID{0000-0002-8117-0612}}

\authorrunning{Chuting Yu et al.}

 \institute{The University of Queensland, Queensland, Australia \\
\email{\{v.yu, g.zuccon, t.leelanupab\}@uq.edu.au}}
\maketitle

\begin{abstract}
\input{sections/abstract.tex}

\vspace{-4pt}
\end{abstract}

\vspace{-24pt}
\section{Introduction}
\label{sec:intro}
\input{sections/introduction.tex}

\vspace{-4pt}
\section{Related Work}
\label{sec:related}
\input{sections/related_work.tex}

\vspace{-8pt}
\section{Experimental Methodology}
\vspace{-6pt}
\label{sec:methodology}
\input{sections/method.tex}

\vspace{-18pt}
\section{Results}
\label{sec:results}
\vspace{-8pt}
\input{sections/results.tex}

\vspace{-10pt}
\section{Conclusion}
\vspace{-8pt}
\label{sec:conclusion}
\input{sections/conclusion.tex}

\bibliographystyle{splncs04}
\bibliography{custom}


\end{document}

%% file: sections/abstract.tex
LLMs are evolving rapidly, with newer models offering stronger capabilities. This suggests that in LLM-based relevance judging, more capable models will achieve higher agreement with human judgements under the same prompt.
%
We challenge this understanding by investigating the behavior of LLM-based relevance judges under backbone evolution. Keeping the prompts fixed, we evaluate a representative single-prompt (UMBRELA) and a rubric-based prompt (EXAM) across sequential model versions of commercial (Gemini, GPT) and open-weight (Qwen, Llama) models.
Overall, we find no consistent evidence that newer versions lead to better relevance judges. Crucially, similar or improved aggregate performance does not imply judgment stability: correct judgements made by an earlier version of an LLM backbone are not necessarily preserved by later versions. We investigate the potential drivers of these regressions. Our findings caution against the assumption that judging prompts designed and validated for one backbone version will perform equivalently or better when the model is updated, even within the same family.

%% file: sections/introduction.tex
\vspace{-7pt}
Large language models (LLMs) are increasingly used as \textit{judges} for automatic evaluation, where an LLM is prompted to assess the quality of model outputs or other artefacts in place of, or in support of, human evaluation \cite{faggioli2023perspectives,macavaney2023one,li2024llms}. Recent work shows that appropriately prompted LLMs can produce relevance judgements that achieve substantial agreement with human assessments \cite{rahmaniJudgingJudgesCollection2025,thomas2024large}. Such approaches offer an attractive alternative to costly human relevance assessment and make it possible to construct or extend evaluation collections at scale.

Prompting is a key component of LLM-based judges, guiding the model’s judging behavior and shaping its outputs. In practice, considerable effort can be required to design, refine, and validate such prompts for a particular evaluation setting. Importantly, a prompt does not operate independently of the model executing it~\cite{cao2024worst,wang2025promptbridge}. Different LLMs may interpret the same instructions differently, exhibit different reasoning behaviors, and ultimately produce different judgements. Prior work has primarily examined this dependency across different LLMs, showing that the same judging framework can yield substantially different assessment effectiveness when paired with different model backbones \cite{farzi2025does}.

LLMs are evolving rapidly, with successive model versions continuously released~\cite{echterhoff2024muscle}. While prompt effectiveness is known to vary across different LLMs, less is known about how it changes across \textit{different versions} of the same LLM family(e.g., how a relevance assessment prompt designed using Gemini 2.5 Flash performs when the backbone is updated to Gemini 3.8 Flash). Redesigning and re-validating a judging prompt each time the underlying backbone is updated is costly and undermines one of the main attractions of LLM-based evaluation. It is therefore tempting to replace an existing backbone with a newer, more capable version from the same family, while retaining the same judging prompt and expecting similar or improved performance.

But does a newer version of the same LLM backbone necessarily make a better judge? In this work, we examine this question by studying LLM-based relevance judges across two TREC datasets under model evolution. We keep the judging prompt fixed and vary only the underlying LLM backbone across successive versions of Gemini, GPT, Qwen, and Llama. We consider two different relevance judging approaches: UMBRELA, which directly prompts an LLM to produce a graded relevance assessment~\cite{upadhyay2024umbrela}, and EXAM, which employs a structured rubric-based assessment framework where a set of questions is first inferred for each query, and then the rubric is applied to a document for grading~\cite{farzi2024exam}. Beyond aggregate performance, we examine whether newer model versions from the same backbone family preserve judgements that their predecessors made correctly. In this context, we define a regression as an instance of capability degradation in which an earlier model version predicts the exact ground-truth relevance label, but a subsequent model fails to do so.
Our investigation addresses three research questions --- \textbf{RQ1:} Do newer LLM versions of the same backbone lead to better relevance judges?; \textbf{RQ2:} Are there regressions when switching to a newer version?; \textbf{RQ3:}What are the key drivers of such regressions?


Our results show that newer LLM versions from the same family do not consistently produce better relevance judgments under a fixed prompt. While aggregate performance improves in some cases, it degrades in others; moreover, improved aggregate performance can mask regressions on previously correct judgements. These regressions are more substantial for the complex EXAM rubric-based protocol than for the simpler UMBRELA protocol. Our analysis further identifies systematic changes in judging behavior behind these regressions, including over-weighting specific aspects, overly strict interpretations of relevance criteria, and superficial keyword matching. 
%
By examining both aggregate performance and instance-level regressions, our study shows that model evolution can alter the behavior of an established LLM judge, cautioning practitioners against assuming that judging prompts designed and validated for one model version will transfer reliably to a newer one, even within the same family. This highlights the need to re-validate judging prompts when their underlying backbones are updated.

%% file: sections/related_work.tex
\vspace{-4pt}
To improve agreement with human relevance judgments, prior work has explored various prompt engineering strategies for LLM-based relevance assessment. Thomas et al.~\cite{thomas2024large} systematically investigated different prompt components, including role-based instructions, additional descriptions of the information need, step-by-step reasoning, aspect-based assessment, and simulated multiple judges. Their results, obtained using early version of OpenAI's GPT, showed substantial variation in agreement with human judgments across prompt configurations, with aspect-based assessment providing the largest improvement. Building on this, Upadhyay et al.~\cite{upadhyay2024umbrela} introduced UMBRELA, an open-source reproduction of Thomas et al.'s approach that employs a structured prompt with explicit relevance criteria to guide LLMs in assigning graded relevance labels. 

However, a persistent challenge in the use of LLM judges is prompt sensitivity: minor modifications in framing, formatting, or instruction order can lead to substantial variation in model outputs~\cite{mizrahi2024state,bellibatlu2026judgesense,sclar2024quantifying}. Arabzadeh\&Clarke~\cite{arabzadeh2025human} systematically investigated prompt sensitivity in LLM-based relevance assessment across binary, graded, and pairwise judging settings. By comparing prompts authored by human experts and LLMs, 
they demonstrated that variations in prompt formulation can substantially affect agreement with human assessments, highlighting prompt design as an important source of variability. 

Prompt formulation, however, is not the only source of variability: the choice of the underlying LLM backbone can also substantially affect relevance assessment performance~\cite{rahmaniJudgingJudgesCollection2025,abbasiantaeb2024can}. Prior studies have shown that, under the same relevance-judging prompts, different LLM backbones can exhibit different levels of agreement with human judgments~\cite{alaofi2026use,farzi2025does}. Farzi\&Dietz~\cite{farzi2025does} for example examined whether UMBRELA generalises across different LLMs, finding comparable performance between GPT-4o and DeepSeek V3 but progressively lower effectiveness with LLaMA-3.3-70B and smaller models.


Beyond the choice of prompt and model backbone, a critical but less studied source of instability arises as these models evolve over time. Successive versions of the same LLM service might in fact exhibit behavioral drift on general generative tasks (e.g., coding or reasoning), though initial studies on this phenomenon have been limited to a narrow set of proprietary models~\cite{chen2023chatgpt}. Such changes can have direct implications when evolving models are used for system evaluation. Pozzobon et al.\cite{pozzobon2023challenges} demonstrated this issue in the context of toxicity evaluation, showing that updates to a commercial evaluation API could change previously obtained results and even alter the relative ranking of evaluated models.


More recently, Dietz et al.~\cite{dietz2025principles} identified \emph{LLM evolution} as a direct threat to the validity and reproducibility of LLM-based relevance assessment. While they hypothesise newer model versions may judge relevance differently and advocate for periodic re-evaluation, this threat has yet to be empirically quantified. In particular, it remains unclear whether newer versions preserve their predecessors' correct judgements, or if they introduce regressions despite overall improvements in capabilities. To address this gap, our research systematically investigates the stability of LLM judges across subsequent versions of the same backbone. 

%% file: sections/method.tex
We systematically evaluate how LLM-based relevance judges behave as their underlying model backbones evolve. To isolate the effect of model evolution, we keep the evaluation prompt fixed and vary only the model version within a model family.  Under this controlled setting, we address three research questions by examining aggregate performance changes (RQ1) and instance-level capability regressions across successive model versions (RQ2 and RQ3).

\textit{\textbf{Dataset and Evaluation Task.}}
We conduct our evaluation on the TREC Deep Learning (DL) Track passage ranking task from 2019 and 2020~\cite{craswell2020overview,craswell2021overview}, using human relevance labels (0-3) as ground truth. The graded judgments allow us to evaluate not only exact agreement with human assessments but also the magnitude of judgment errors. 

\vspace{2pt}


\textit{\textbf{Prompting Frameworks.}} We evaluate the models using two structurally distinct relevance assessment prompts. \textbf{UMBRELA} is a zero-shot relevance assessment framework that employs DNA (Descriptive, Narrative, and Aspects) prompting~\cite{upadhyay2024umbrela}. The descriptive and narrative components establish the relevance criteria and the query--passage context, while the aspects component provides a step-by-step procedure that guides the LLM in assessing relevance and assigning a graded relevance label.
\textbf{EXAM} is a two-step rubric-based relevance assessment framework~\cite{farzi2024exam}. In the first step, the LLM generates a set of exam questions for each query, forming a query-specific rubric. In the second step, the LLM assesses the answerability of each exam question using the passage as context, assigning a rating from 0 to 5. These question-level assessments are then used to derive the passage-level relevance grade on the 0–3 TREC relevance scale.

For EXAM, we use the earliest model version within each family to generate the query-specific questions and fix these questions across all subsequent versions in that family, ensuring that each model is evaluated using the same rubric. This is so that any observed regressions are strictly attributable to the model's assessment behavior: if both the rubric generator and grader evolved simultaneously, shifts in question generation would conflate our measurement of the model's core grading stability.
For both UMBRELA and EXAM, we build on their publicly available GitHub implementations and extend them to support the native APIs of the commercial models evaluated in our experiments. This allows us to preserve model-specific capabilities, including reasoning where supported. 

\begin{figure}[t]
    \centering
    \includegraphics[width=0.98\textwidth]{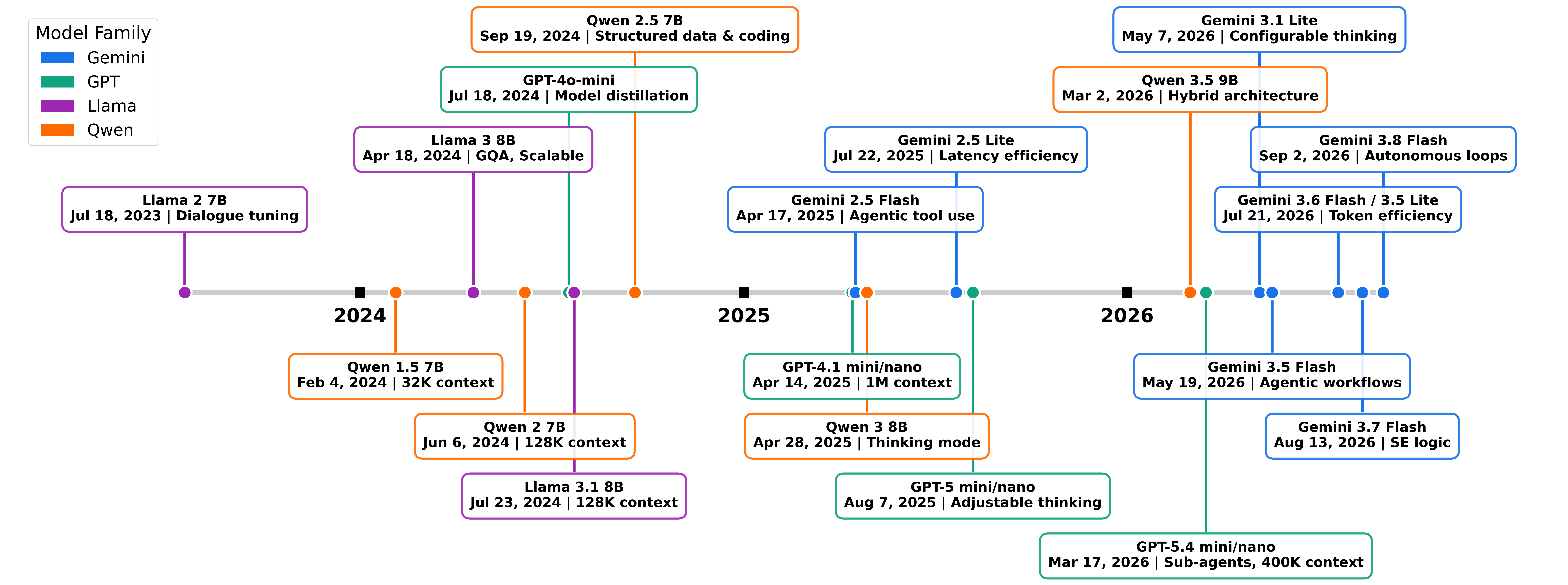}
    \vspace{-12pt}
    \caption{Release dates and key innovations of the evaluated LLMs.
    \vspace{-18pt}
    }
    \label{fig:model_timeline}
\end{figure}

\vspace{2pt}
\textit{\textbf{LLM Backbones.}}
We study model evolution across four LLM families. Gemini and GPT are 
commercial, closed-source LLMs, while Qwen and Llama provide 
open-weight counterparts. For each family, we select successive releases 
within a consistent model lineage, enabling us to examine cross-version 
changes in relevance-judging behavior while keeping the prompting framework 
and evaluation setting fixed. Model families, versions and lineages evaluated are visualised in Figure~\ref{fig:model_timeline}; exact versions are listed in \url{https://github.com/ielab/judge-backbone-evolution}, which also contains code and results of experiments. Commercial models were accessed via their native APIs. For Qwen and Llama, successive generations do not always provide instruction-tuned checkpoints at identical parameter scales. We therefore select the closest available model sizes across generations (7--9B) to minimize differences in model capacity while preserving the chronological model lineage. For all models, $temperature=0$.





\vspace{2pt}
\textit{\textbf{Aggregate Performance and Instance-Level Changes.}}
We analyze relevance-judging behavior across backbone versions at two levels:
aggregate performance and instance-level changes. At the aggregate level,
we evaluate each model version against human relevance judgments to determine whether judging performance improves or degrades as the model  (RQ1). At the instance level, we examine how individual query--document judgments change across successive model versions, with a particular focus on regressions (RQ2 and RQ3).

For agreement with human judgments, we use three metrics: Exact Match (EM),
which measures the percentage of predictions that exactly match the human
relevance grade; Mean Absolute Error (MAE), which captures the magnitude of
grading errors; and Pearson correlation ($r$), which measures the linear
association between model-predicted and human relevance grades.

To characterize changes across model  at instance level, we focus on
\emph{regressions}. For a given query--document pair $(q,d)$ with human
relevance label $y$, we define a regression as an instance for which the
baseline model $M_{\mathrm{base}}$ produces the correct relevance grade,
while a subsequent model $M_{\mathrm{new}}$ does not:
\begin{equation}
\vspace{-8pt}
M_{\mathrm{base}}(q,d) = y
\quad \land \quad
M_{\mathrm{new}}(q,d) \neq y. \label{eq_regression}
\end{equation}

%% file: sections/results.tex
\subsection{RQ1: Do newer LLM versions lead to better relevance judges?}
\label{sec_rq1}
\vspace{-4pt}
We first examine how aggregate relevance-judging performance changes across successive model versions. Table~\ref{tab:combined_results_all} reports the results across model families, datasets, and prompting frameworks.

\input{tables/combined_results_table_merged}

 Overall, we find no consistent evidence that newer model versions lead to better relevance judges: while some model families show relatively steady gains across successive releases, others exhibit substantial fluctuations across versions.

Gemini Flash shows an initial improvement followed by saturation or mild decline, rather than continuous gains across releases. On DL2019, under UMBRELA, performance peaks at version 3.6 (48.28\% EM, 0.597 MAE) before declining in 3.7 and 3.8. Under EXAM, performance is more stable, increasing from 44.81\% EM at version 2.5 to approximately 52\% across the 3.x versions.

In contrast, Gemini Flash Lite shows an overall trend of improvement across successive releases, with EM consistently increasing and MAE generally decreasing from version 2.5 through 3.5 across both prompts and datasets. However, the effect of the same model version update can differ substantially across judging frameworks. For GPT Mini, the transition from GPT-4o Mini to GPT-5 Mini reduces UMBRELA EM (DL 2019: $55.80\% \rightarrow 35.63\%$; DL 2020 $60.26\% \rightarrow 38.71\%$), 
while EXAM EM increases (DL 2019: $38.95\% \rightarrow 45.06\%$; DL 2020 $50.40\% \rightarrow 51.44\%$).
This contrast suggests that the impact of a model update may depend on the judging framework in which it is evaluated.

Among the open-weight families, Qwen exhibits substantial performance fluctuations across successive versions. Under UMBRELA, performance improves from Qwen1.5-7B to Qwen2.5-7B, which achieves the highest EM on both DL 2019 (58.12\%) and DL 2020 (59.45\%), before declining in later versions. Under EXAM, EM drops sharply from Qwen2.5-7B to Qwen3-8B (DL 2019: $47.87\% \rightarrow 15.15\%$; DL 2020: $55.32\% \rightarrow 19.30\%$), before recovering with Qwen3.5-9B (56.43\% and 69.00\%, respectively). By comparison, Llama shows a more consistent improvement across successive versions. EM increases from Llama2-7B through Llama3-8B to Llama3.1-8B under both UMBRELA and EXAM on both datasets, accompanied by generally lower MAE and higher Pearson correlation. 

Overall, the results reveal that changes across model versions are not always consistent across evaluation metrics, and that these changes can vary with the judging framework. Improvements in EM do not necessarily coincide with lower MAE or higher Pearson $r$; for instance, UMBRELA generally yields stronger Pearson $r$ and lower MAE for Gemini Flash, whereas EXAM achieves higher EM for several versions. This suggests that model evolution can affect different aspects of relevance-judging performance differently, and that the effect of a model update should be considered in conjunction with the judging framework in which it is used. Consequently, the performance of a fixed judging framework may need to be re-evaluated as the underlying model evolves. However, these aggregate metrics describe only overall performance and do not reveal whether newer model versions preserve the individual judgments that their predecessors made correctly. We examine this question through instance-level regression analysis in RQ2.


\textbf{Stability Analysis.} While we evaluate all models using greedy decoding ($temperature=0$) to isolate backbone differences, commercial API endpoints can introduce generation variance due to hidden backend operations such as speculative decoding, mixture-of-experts routing, or on-the-fly request sharding. To verify empirically that our findings stem from genuine capability shifts rather than transient API noise, we conducted a stability analysis. Given the prohibitive computational cost of running multiple inferences across all studied families, we selected the Gemini Flash series as a representative testbed for commercial APIs, operating under the assumption that open-weight models executed locally exhibit strict determinism at temperature zero. By running three independent repeats across the DL 2019 assessments for each Gemini model, we found that aggregate performance metrics are highly stable (EM's standard deviation $<0.1\%$; MAE and $r$ $<0.003$). This confirms that the cross-version changes we observe are consistent characteristics of the evolving backbones.

\vspace{-8pt}
\subsection{RQ2: Are there regressions when switching to a newer version?}
\vspace{-4pt}

We operationalise a regression as a strict loss of prior capability between model versions. Specifically, for a given query-document pair, a regression occurs if and only if the baseline model ($M_{base}$) produces an assessment that is exactly concordant with the human ground-truth relevance label ($y$), whereas a subsequent model ($M_{new}$) diverges from it (see Equation~\ref{eq_regression}). 



\begin{figure*}[t]
    \centering
    \begin{subfigure}[b]{0.48\textwidth}
        \centering
        \includegraphics[width=\textwidth]{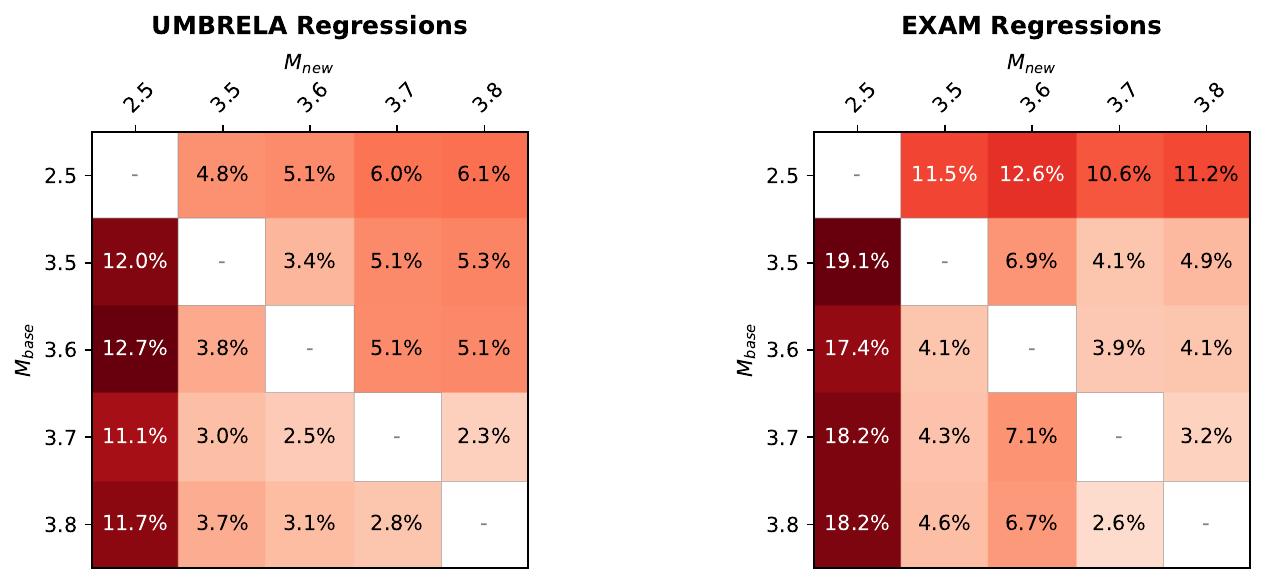}
        \caption{Gemini Flash}
        \label{fig:heat_gemini_flash}
    \end{subfigure}
    \hfill
    \begin{subfigure}[b]{0.48\textwidth}
        \centering
        \includegraphics[width=\textwidth]{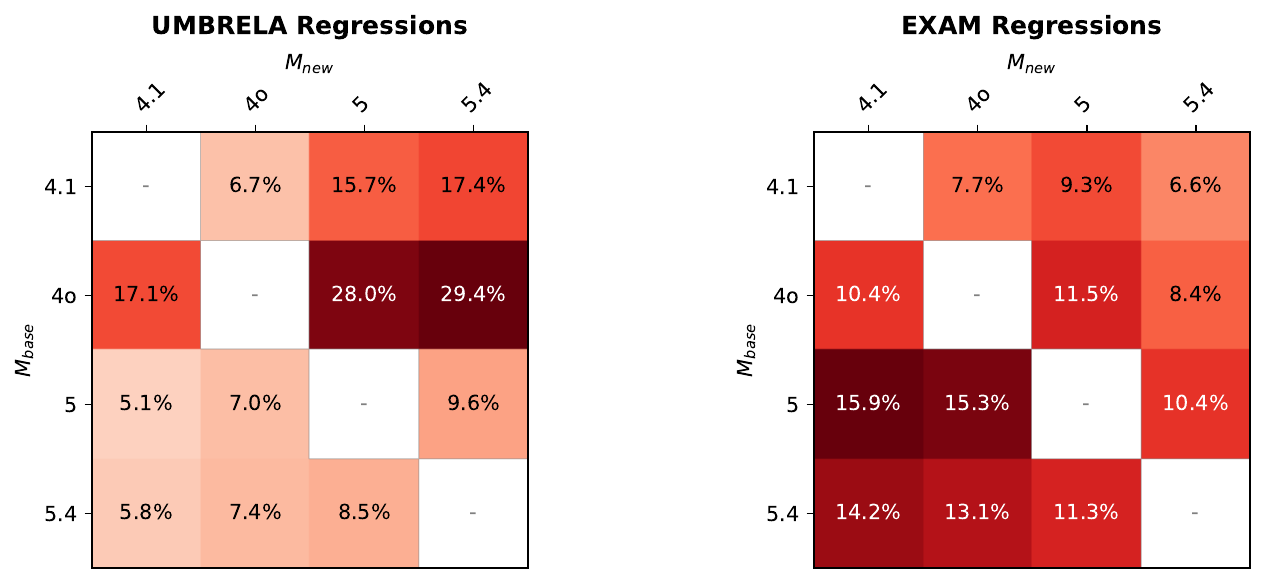}
        \caption{GPT Mini}
        \label{fig:heat_gpt_mini}
    \end{subfigure}
    
    
    \begin{subfigure}[b]{0.48\textwidth}
        \centering
        \includegraphics[width=\textwidth]{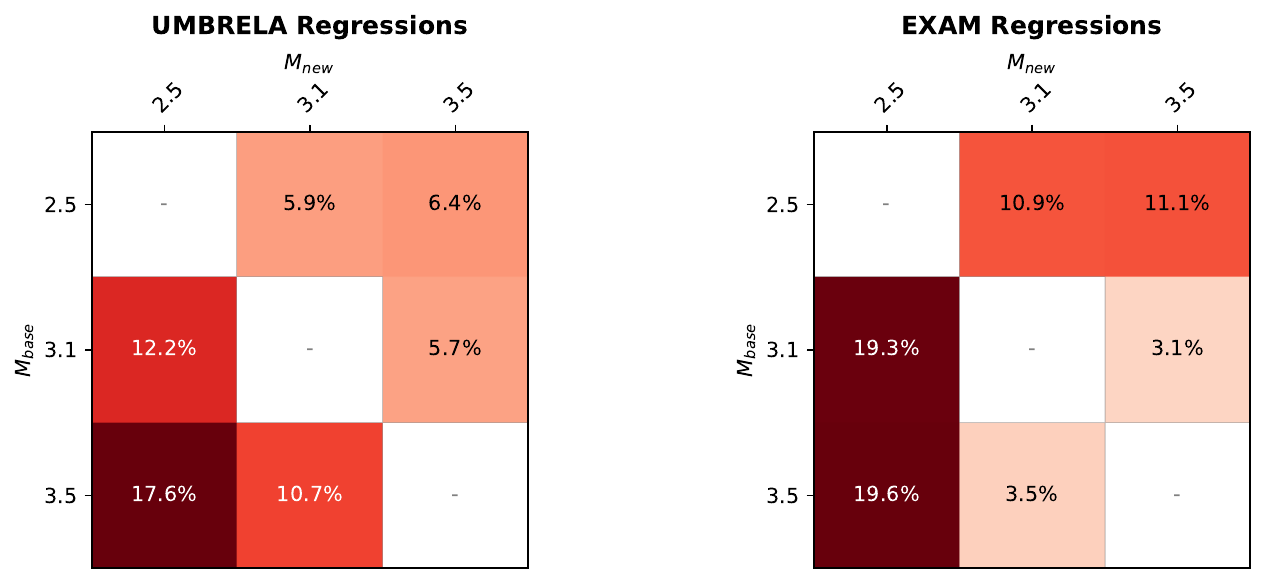}
        \caption{Gemini Flash Lite}
        \label{fig:heat_gemini_lite}
    \end{subfigure}
    \hfill
    \begin{subfigure}[b]{0.48\textwidth}
        \centering
        \includegraphics[width=\textwidth]{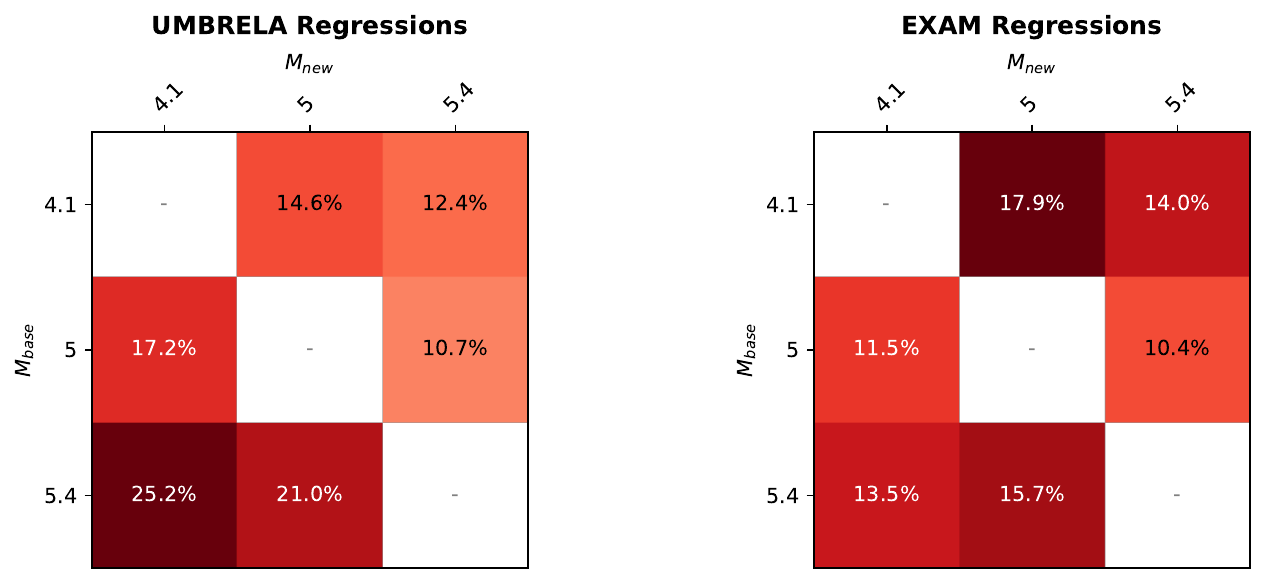}
        \caption{GPT Nano}
        \label{fig:heat_gpt_nano}
    \end{subfigure}
    
    
    \begin{subfigure}[b]{0.48\textwidth}
        \centering
        \includegraphics[width=\textwidth]{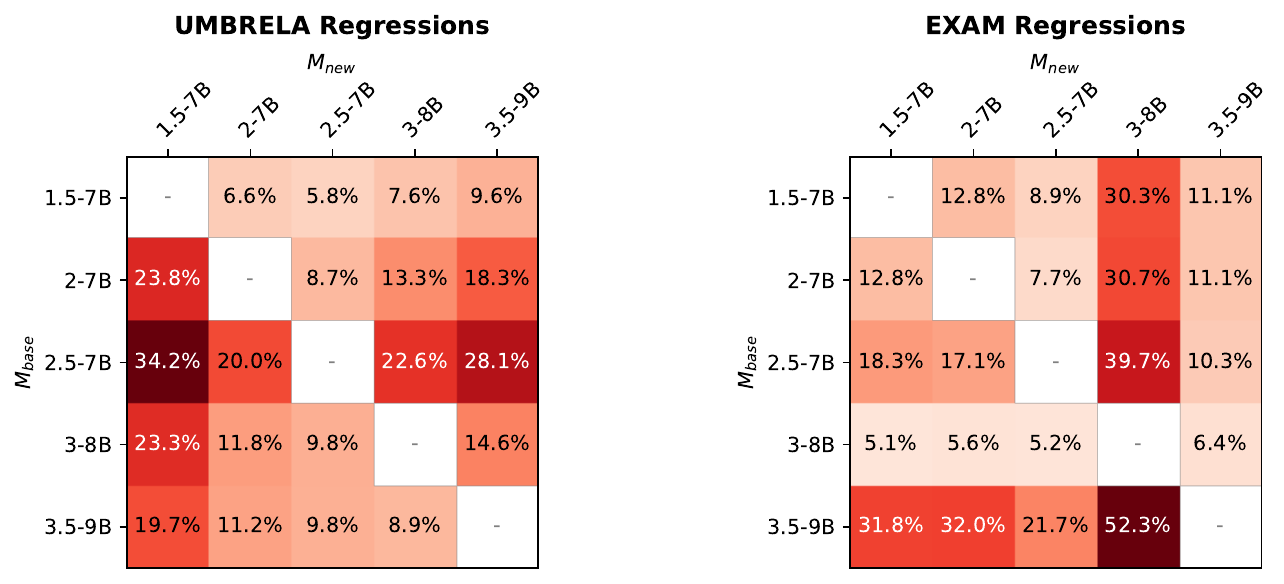}
        \caption{Qwen}
        \label{fig:heat_qwen}
    \end{subfigure}
    \hfill
    \begin{subfigure}[b]{0.48\textwidth}
        \centering
        \includegraphics[width=\textwidth]{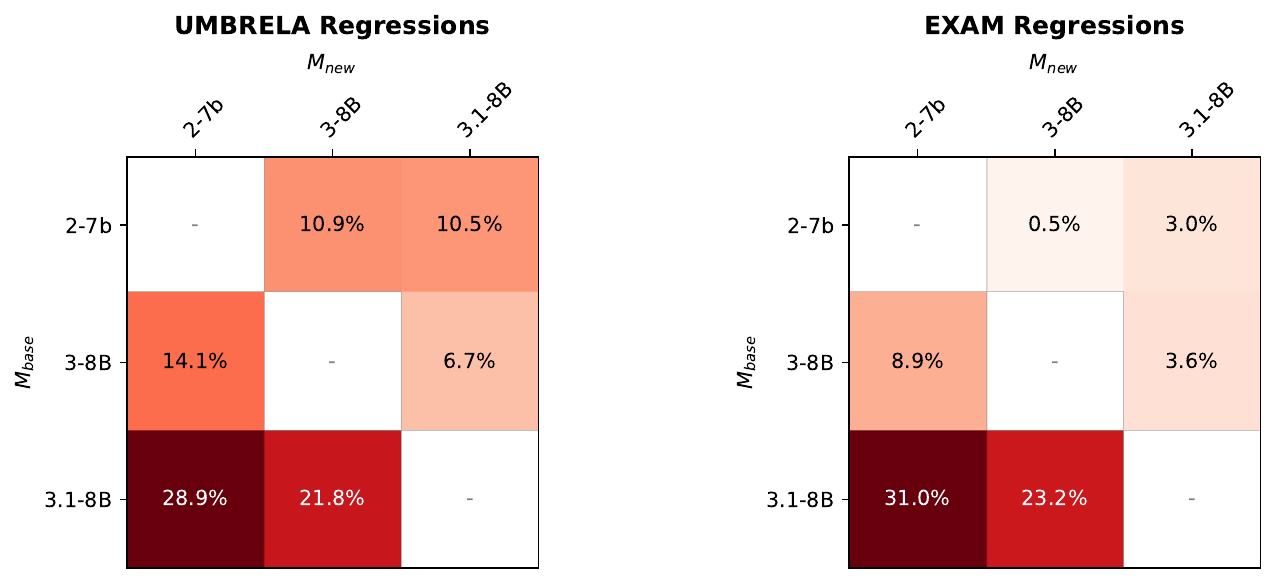}
        \caption{Llama}
        \label{fig:heat_llama}
    \end{subfigure}
    \vspace{-10pt}
    \caption{Instance-level regression rates across evaluated model families. Each figure presents the regressions (combined across TREC DL 2019 and 2020) for the zero-shot UMBRELA prompt (left) and the rubric-based EXAM prompt (right). Rows indicate the baseline model ($M_{base}$); columns the newer model ($M_{new}$). \vspace{-20pt}}
    \label{fig:all_heatmaps}
\end{figure*}

The regression analysis on the combined TREC DL 2019 and 2020 datasets across the six studied model families is visualised in Figure \ref{fig:all_heatmaps}\footnote{Early Llama versions often failed to output valid EXAM scores, artificially suppressing their regression rates (Section~\ref{sec_rq1} -- we report how we handled those in our GitHub repository). We thus interpret their EXAM stability with caution.}. The analysis uncovers several critical insights regarding the evolution of LLMs as relevance judges under both zero-shot and rubric-based evaluation frameworks.

The aggregate-level analysis reveals substantial churn across all evaluated model families (RQ1). Even when newer models improve in overall EM, these aggregate gains do not necessarily reflect the preservation of previously correct judgments. Under the UMBRELA prompt, for instance, while Gemini 3.5 Flash Lite sensibly improves upon 3.1 in EM (Table~\ref{tab:combined_results_all}), 5.7\% of instances correctly assessed by the older model were incorrectly graded by the newer one.
This trend is also visible in open-weight families; transitioning from Llama 3 to 3.1 incurs a 6.7\% regression rate despite a near-double improvement in EM. This indicates that the widely assumed ``drop-in'' replacement paradigm\footnote{i.e., the common practice of substituting an older LLM with its newer version by simply updating the API endpoint or model identifier in an existing pipeline, relying on the assumption that newer models are strictly superior upgrades that will preserve or enhance previous capabilities without necessitating any prompt re-engineering.} is fundamentally flawed; new versions shift the underlying decision boundaries and prompt alignment rather than strictly subsuming past capabilities.

This volatility is further magnified during major architectural or generational shifts, which often induce extreme instability. The transition from GPT-4o Mini to GPT-5 Mini resulted in a severe performance collapse for the UMBRELA prompt, losing over a quarter (28.0\%) of the correct judgments made by its predecessor. A similar disruption occurs in the Qwen family, where upgrading from Qwen 2.5 to Qwen 3.0 incurs a 22.6\% regression rate. These severe capability losses highlight that major model overhauls can severely disrupt prompt-specific performance and break established evaluation pipelines.

Conversely, minor, within-generation updates can lead to greater stability. Among all the families studied, the Gemini Flash family provides a quite unique perspective on this due to the rapid succession of development of the 3.5 through 3.8 models, which were all released within a compressed 3.5-month window between May and September 2026. As the underlying model iteratively matured with token efficiency tweaks and reasoning refinements, the regression rates gradually declined. The final transition from Gemini 3.7 Flash to 3.8 Flash exhibits a remarkably low regression rate of 2.3\% under UMBRELA, suggesting that as this architectural backbone was fine-tuned, its grading behaviour stabilises.

When shifting our analysis to EXAM (Figure \ref{fig:all_heatmaps}, right subplots), we find that this more complex, rubric-based evaluation framework amplifies the effects of backbone evolution in both directions. 
For minor model updates, EXAM significantly exacerbates volatility at the query-document pair level. We hypothesise that by forcing the model down a rigid, multi-step logical pathway, any minor shifting of weights can easily change one step of the rubric, causing the entire assessment to change -- and this is despite we kept the actual rubric generation step fixed\footnote{i.e., the rubric was generated once and shared across all versions of the same family.}. Consequently, the EXAM prompt induces much higher regression rates across version transitions than the UMBRELA prompt for nearly all model families. When upgrading from Gemini 2.5 Flash to 3.5, regressions jump from 4.8\% for UMBRELA to 11.5\% for EXAM. 

However, during severe generational degradation, the structure of the EXAM prompt provides a ``buffering'' effect. 
For example, the 28.0\% UMBRELA regression from GPT-4o Mini to GPT-5 Mini fell to just 11.5\% under EXAM.
We hypothesise that this buffering occurs because the structured rubric provides external scaffolding: even when a model's zero-shot alignment is disrupted, its instruction-following capabilities can still ground the judgment step-by-step against explicit rules, effectively mitigating the capability loss.


Finally, the regression heatmaps reveal an asymmetry in capability shifts, which is particularly relevant given the economic incentive to downgrade to older models as their API costs decrease following the release of newer versions.\footnote{This is because LLM providers often discount API costs for the use of older models upon the release of a newer model. For example, following the rollout of GPT-4 variants throughout 2023 and 2024, OpenAI dropped the API pricing of the older GPT-3.5 Turbo three times within a single year.} When a newer version represents a successful upgrade, reverting to its predecessor can incur substantially higher regression than the forward transition\footnote{Here we override regression with $M_{\mathrm{new}}(q,d) = y
\quad \land \quad
M_{\mathrm{old}}(q,d) \neq y.$} than the forward upgrade. For example, upgrading from Llama 3 to 3.1 incurs a 6.7\% regression under UMBRELA, but downgrading from 3.1 back to 3 loses 21.8\% of the newer model's correct assessments; other model families show similar patterns of downgrade losses. 
This highlights that while older models may become significantly cheaper as new ones are released, relying on them as budget alternatives to newer models sacrifices a disproportionately large fraction of the correct relevance assessments achieved by the newer model.

\vspace{-6pt}
\subsection{RQ3: What are the drivers beyond the regressions?}
\vspace{-4pt}

To understand the qualitative nature of performance regressions, we conducted a meta-evaluation of the reasoning trajectories produced by the baseline and subsequent models for both the UMBRELA and EXAM prompts. While our overarching goal is to understand regression drivers generally, the substantial computational cost of this analysis, combined with space constraints, necessitated focusing on a single model family. We selected Gemini Flash because its rapid succession of distinct versions provides a particularly granular testbed for tracing how regression behaviour evolves across successive model updates. 

For any given pair of baseline model $M_{base}$ and subsequent model $M_{new}$, we first isolated the subset of query-document pairs that resulted in a strict regression. 
For each regressed instance, we extracted the explicit reasoning trajectories generated by both models prior to outputting their final relevance judgments. 
We then used Gemini 3.8 Flash as an expert meta-evaluator to diagnose why the newer model failed. For each regression, the meta-evaluator was provided with the original query and document text, the ground-truth human relevance score, and the scores and reasoning trajectories produced by both $M_{base}$ and $M_{new}$. The meta-evaluator was prompted to conduct a comparative analysis of the two reasoning trajectories and identify exactly where $M_{new}$ diverged or failed. It was instructed to output a concise classification of the primary failure mode and a brief qualitative justification detailing the flaw in $M_{new}$'s reasoning compared to the successful reasoning of $M_{base}$. Classification categories were manually reviewed and aggregated when redundant. We limited our analysis on TREC DL 2019 for budget reasons,  
processing 4,933 UMBRELA regressions and 6,854 EXAM regressions. To validate the meta-evaluator, we manually audited a sample of 50 regressions and found the LLM’s categorisation to be consistent with human interpretation.




Our analysis (Table \ref{tab:regression_thinking_analysis_comparison}) reveals capability degradation does not stem from uniform reasoning deterioration. Instead, failures cluster into distinct categories heavily influenced by the underlying prompt (i.e. whether UMBRELA or EXAM).

\begin{table*}[t!]
  \centering
  \small 
  \resizebox{0.92\textwidth}{!}{%
  \begin{tabular}{p{80pt} p{8cm} rr rr}
    \toprule
    & & \multicolumn{2}{c}{\textbf{UMBRELA}} & \multicolumn{2}{c}{\textbf{EXAM}} \\
    \cmidrule(lr){3-4} \cmidrule(lr){5-6}
    \textbf{Failure Mode} & \textbf{Description} & \textbf{Count} & \textbf{\%} & \textbf{Count} & \textbf{\%} \\
    \midrule
    Over-Weighted Specific Aspect & $M_{new}$ fixated too heavily on a single keyword, constraint, or tangential topic in the document, allowing it to overshadow the broader semantic relevance of the text & 1543 & 31.3\% & 72 & 1.1\% \\
    Overly Strict Interpretation & $M_{new}$ applied the grading criteria too rigidly, e.g., penalizing a document completely because it lacked one minor detail, whereas $M_{base}$ correctly labelled it as highly relevant overall & 684 & 13.9\% & 563 & 8.2\% \\
    Superficial Keyword Matching & $M_{new}$ relied on shallow lexical overlap (exact word match) rather than deep semantic understanding & 314 & 6.4\% & 204 & 3.0\% \\
    Formatting Issue & $M_{new}$'s output broke the required rubric schema (e.g., failing to format the score correctly), leading to a parsing failure or forced incorrect score. & 225 & 4.6\% & 0 & 0.0\% \\
    Hallucination & $M_{new}$ hallucinated a requirement that wasn't actually requested in the original query, or assumed the presence of content that wasn't actually in the document. & 208 & 4.2\% & 541 & 7.9\% \\
    Overcrediting Tangential Relevance & $M_{new}$ gave too high of a score to a document that only passingly or incidentally mentioned the topic, failing to recognise that the document wasn't dedicated to answering the query & 113 & 2.3\% & 973 & 14.2\% \\
    Missing Reasoning & $M_{new}$ failed to generate a sufficient reasoning trajectory or skipped the thinking step entirely before outputting a score, leading to a poorly grounded judgment & 99 & 2.0\% & 357 & 5.2\% \\
    Ignored Nuance & $M_{new}$ missed the contextual, temporal, or highly specific nuanced meaning of the query, treating it instead as a generic, broad search & 96 & 1.9\% & 147 & 2.1\% \\
    Overestimated Completeness & $M_{new}$ assumed the document provided a fully exhaustive answer to the query when it only provided a partial one, leading to an inflated score & 52 & 1.1\% & 0 & 0.0\% \\
    Missed Key Information & $M_{new}$ completely overlooked or failed to extract a critical piece of information present in the document that $M_{base}$ successfully identified & 42 & 0.9\% & 0 & 0.0\% \\
    Overly Strict Relevance Threshold & $M_{new}$ scored a marginally relevant document as completely irrelevant (0), applying an overly strict threshold and missing subtle semantic connections that $M_{base}$ successfully caught & 39 & 0.8\% & 2284 & 33.3\% \\
    Overcrediting Topical Overlap & $M_{new}$ gave a high score because the document was about the exact same broad topic as the query, even though it didn't actually answer the specific question being asked & 39 & 0.8\% & 64 & 0.9\% \\
    Misinterpretation Of Query/Doc & $M_{new}$ fundamentally misunderstood the core intent of the query or the primary thesis of the document & 31 & 0.6\% & 16 & 0.2\% \\
    Overcrediting Broad Topical Overlap & $M_{new}$ rewarded a document simply for sharing the same general domain/theme as the query without addressing the specific information need & 29 & 0.6\% & 0 & 0.0\% \\
    Score/Reasoning Inconsistency & $M_{new}$ generated a sound reasoning trajectory that accurately evaluated the document, but then inexplicably output a final numeric score that directly contradicted its own written reasoning & 28 & 0.6\% & 1257 & 18.3\% \\
    \midrule
    \textbf{Total (Selected Categories)} & & \textbf{3542} & \textbf{71.8\%} & \textbf{6478} & \textbf{94.5\%} \\
    \bottomrule
  \end{tabular}%
  }
  \caption{Description and prevalence of the primary  failure modes causing capability degradation across the Gemini Flash family. Reported are categories with prevalence $\ge 0.6\%$ for either UMBRELA or EXAM on DL 2019. 
  \vspace{-34pt}
  }
  \label{tab:regression_thinking_analysis_comparison}
\end{table*}

Under UMBRELA, regressions are predominantly driven by semantic hyper-fixation. The single largest driver of regression, accounting for nearly a third of all UMBRELA failures (31.3\%), is the \textit{Over-Weighted Specific Aspect} mode. In these instances, the newer model fixates heavily on a single keyword, an isolated constraint, or a tangential topic present in the document. This causes the model to allow a minor or peripheral element to entirely overshadow the broader semantic relevance of the text, a capability that the baseline models successfully demonstrated. This suggests that as models evolve, they may become overly sensitive to specific tokens at the expense of holistic document comprehension. 

This loss of holistic understanding under UMBRELA is further evidenced by a reversion to \textit{Superficial Keyword Matching} (6.4\%), where the newer model erroneously equates exact lexical overlap with relevance, and \textit{Overly Strict Interpretation} (13.9\%), where the model penalises highly relevant documents for lacking a minor, implicit detail. 

Conversely, the failures observed under the rubric-based EXAM prompt are characterised by an inflexible application of the grading rules and logical contradictions. The most prevalent failure mode for EXAM is the application of an \textit{Overly Strict Relevance Threshold} (33.3\%). In these cases, the rigid rubric forces the newer model into excessively harsh judgments, scoring marginally relevant documents as completely irrelevant (0) and missing subtle semantic connections that the baseline models successfully caught. EXAM also suffers heavily from \textit{Overcrediting Tangential Relevance} (14.2\%), where the newer model inflates the score of a document simply because it passes a rubric checkpoint by incidentally mentioning a topic, failing to recognise that the document does not actually answer the query~\cite{yu2026llm}.

Most critically, the EXAM prompt induces a high rate of \textit{Score/Reasoning Inconsistency} (18.3\%). In these instances, the newer model generates a sound reasoning trajectory that accurately evaluates the document, but inexplicably outputs a final numeric score that directly contradicts its own reasoning. This finding provides direct, qualitative evidence validating our hypothesis in RQ2: the highly-constrained, multi-step inference process of EXAM introduces severe fragility. Nearly a fifth of all EXAM regressions occur not because the model misunderstood the text, but because it failed the rubric's aggregation step. 

Finally, a persistent tail of regressions across both prompts is driven by generative instabilities rather than reasoning flaws. \textit{Hallucinations} are prevalent in both frameworks (4.2\% in UMBRELA, 7.9\% in EXAM), causing the model to either invent requirements not present in the original query or assume the existence of content not present in the document.

%% file: tables/combined_results_table_merged.tex
\begin{table}[t]
\centering
\caption{Exact Match (\textbf{EM}) percentage, Mean Absolute Error (\textbf{MAE}), and Pearson $\mathbf{r}$ correlation, obtained by different releases of the studied backbones across the UMBRELA and the EXAM prompts. 
Best result per family is bolded.}
\label{tab:combined_results_all}
\resizebox{0.95\textwidth}{!}{%
\begin{tabular}{@{}llcccccccccccc@{}}
\toprule
 &  & \multicolumn{6}{c}{\textbf{TREC DL 2019}} & \multicolumn{6}{c}{\textbf{TREC DL 2020}} \\ \cmidrule(lr){3-8} \cmidrule(l){9-14} 
 &  & \multicolumn{3}{c}{\textbf{UMBRELA}} & \multicolumn{3}{c}{\textbf{EXAM}} & \multicolumn{3}{c}{\textbf{UMBRELA}} & \multicolumn{3}{c}{\textbf{EXAM}} \\ \cmidrule(lr){3-5} \cmidrule(lr){6-8} \cmidrule(lr){9-11} \cmidrule(l){12-14} 
\textbf{Backbone} & \textbf{Version} & \textbf{EM} & \textbf{MAE} & $\mathbf{r}$ & \textbf{EM} & \textbf{MAE} & $\mathbf{r}$ & \textbf{EM} & \textbf{MAE} & $\mathbf{r}$ & \textbf{EM} & \textbf{MAE} & $\mathbf{r}$ \\ \midrule
 & 2.5 & 42.70 & 0.704 & 0.580 & 44.81 & 0.730 & 0.324 & 44.23 & 0.692 & 0.551 & 52.88 & 0.627 & \textbf{0.457} \\
Gemini & 3.5 & 47.61 & 0.606 & 0.601 & 52.22 & 0.718 & 0.268 & 53.31 & 0.549 & \textbf{0.602} & 54.12 & 0.638 & 0.435 \\
Flash & 3.6 & \textbf{48.28} & \textbf{0.597} & \textbf{0.601} & 49.78 & 0.716 & 0.262 & \textbf{53.59} & \textbf{0.541} & 0.594 & 59.83 & 0.575 & 0.210 \\
 & 3.7 & 45.01 & 0.621 & 0.600 & \textbf{52.58} & \textbf{0.672} & \textbf{0.335} & 51.64 & 0.553 & 0.596 & \textbf{62.63} & \textbf{0.509} & 0.298 \\
 & 3.8 & 46.65 & 0.604 & 0.594 & 52.31 & 0.693 & 0.322 & 51.23 & 0.551 & 0.585 & 61.25 & 0.518 & 0.317 \\ \midrule
 Gemini & 2.5 & 45.12 & 0.672 & 0.587 & 44.57 & 0.728 & \textbf{0.321} & 48.93 & 0.726 & 0.564 & 53.57 & 0.606 & \textbf{0.322} \\
 Flash & 3.1 & 46.80 & 0.643 & 0.591 & 52.10 & \textbf{0.720} & 0.255 & 58.84 & 0.538 & \textbf{0.601} & 62.78 & 0.528 & 0.212 \\
 Lite & 3.5 & \textbf{54.10} & \textbf{0.552} & \textbf{0.600} & \textbf{52.66} & 0.721 & 0.233 & \textbf{61.96} & \textbf{0.461} & 0.583 & \textbf{63.31} & \textbf{0.516} & 0.222 \\ \midrule
 & 4.1 & 45.79 & 0.654 & 0.583 & 37.16 & 1.019 & 0.429 & 49.57 & 0.630 & \textbf{0.560} & 46.51 & 0.843 & 0.467 \\
GPT & 4o & \textbf{55.80} & \textbf{0.556} & \textbf{0.609} & 38.95 & 0.996 & 0.420 & \textbf{60.26} & \textbf{0.514} & 0.550 & 50.40 & 0.768 & \textbf{0.472} \\
Mini & 5 & 35.63 & 0.821 & 0.557 & \textbf{45.06} & \textbf{0.798} & 0.366 & 38.71 & 0.795 & 0.544 & \textbf{51.44} & \textbf{0.703} & 0.424 \\
 & 5.4 & 35.13 & 0.825 & 0.578 & 44.52 & 0.813 & \textbf{0.449} & 37.16 & 0.822 & 0.536 & 47.96 & 0.785 & 0.428 \\ \midrule
GPT & 4.1 & 39.44 & 0.748 & 0.402 & 47.83 & 0.715 & 0.353 & 40.74 & 0.731 & 0.392 & 52.88 & 0.686 & 0.252 \\
Nano & 5 & 41.02 & 0.741 & \textbf{0.564} & 44.24 & 0.803 & 0.381 & 44.15 & 0.702 & \textbf{0.558} & 48.41 & 0.726 & 0.451 \\
 & 5.4 & \textbf{51.09} & \textbf{0.668} & 0.560 & \textbf{49.30} & \textbf{0.673} & \textbf{0.422} & \textbf{54.48} & \textbf{0.687} & 0.479 & \textbf{53.18} & \textbf{0.613} & \textbf{0.470} \\ \midrule
\multirow{5}{*}{Qwen} 
 & 1.5-7B & 26.92 & 1.567 & 0.304 & 39.90 & 0.920 & 0.347 & 33.22 & 1.558 & 0.280 & 44.83 & 0.866 & 0.388 \\
 & 2-7B & 45.85 & 0.806 & 0.482 & 39.81 & 0.918 & 0.302 & 49.02 & 0.768 & 0.441 & 44.79 & 0.839 & 0.389 \\
 & 2.5-7B & \textbf{58.12} & \textbf{0.553} & 0.535 & 47.87 & 0.773 & 0.369 & \textbf{59.45} & \textbf{0.535} & 0.445 & 55.32 & 0.669 & 0.396 \\
 & 3-8B & 42.76 & 0.749 & \textbf{0.547} & 15.15 & 1.785 & 0.243 & 48.76 & 0.685 & \textbf{0.532} & 19.30 & 1.813 & 0.318 \\
 & 3.5-9B & 36.09 & 0.871 & 0.544 & \textbf{56.43} & \textbf{0.691} & \textbf{0.371} & 44.03 & 0.790 & 0.523 & \textbf{69.00} & \textbf{0.445} & \textbf{0.427} \\ \midrule
\multirow{3}{*}{Llama} 
 & 2-7b & 21.97 & 1.253 & 0.240 & 7.60 & 2.208 & -0.002 & 18.72 & 1.397 & 0.223 & 5.68 & 2.479 & 0.013 \\
 & 3-8B & 22.95 & 1.186 & 0.425 & 14.13 & 1.758 & \textbf{0.299} & 23.77 & 1.241 & 0.399 & 14.14 & 1.926 & 0.274 \\
 & 3.1-8B & \textbf{35.93} & \textbf{1.125} & \textbf{0.455} & \textbf{31.06} & \textbf{1.192} & 0.290 & \textbf{40.71} & \textbf{1.130} & \textbf{0.411} & \textbf{33.76} & \textbf{1.204} & \textbf{0.320} \\ \bottomrule
\end{tabular}%
\vspace{-58pt}
}
\end{table}

%% file: sections/conclusion.tex
Newer LLMs are often assumed to be strictly superior, ``drop-in'' replacements for their predecessors. 
While recent commentary challenges this theoretically~\cite{dietz2025principles}, empirical evidence  remains scarce. In this work, we longitudinally evaluated LLM-based relevance judges across representative model families. 
We find aggregate accuracy gains consistently mask severe instance-level volatility, where new versions routinely overturn correct assessments. Crucially, prompt structure dictates regression severity: the rigid EXAM rubric exacerbated volatility during minor updates but buffered severe generational degradation. 
Our meta-evaluation revealed that capability degradation stems not from uniform reasoning loss, but clusters into prompt-specific categories.
Under UMBRELA, models exhibited semantic hyper-fixation; under EXAM, regressions stemmed from scoring inconsistencies, where final scores contradicted the models'  sound reasoning.

These findings carry significant practical implications. 
High cross-version volatility means IR evaluation pipelines cannot be ``set and forget'': every backbone update requires re-validating prompt-model alignment against human baselines. 
Furthermore, asymmetric regression rates warn against using older, discounted models as budget alternatives, which disproportionately sacrifices capabilities.

Our study has several limitations that present avenues for future work. 
First, our evaluation is restricted to TREC DL; future studies should verify if these behaviours generalise across other tasks and multi-lingual corpora.
Second, we evaluated only UMBRELA and EXAM, leaving alternative frameworks like ensemble judges unexplored. In addition, we kept EXAM's question generator static to isolate the grader's evolution; future work should assess how upgrading the generator alongside the grader impacts stability. 
Finally, to address our findings, future work should explore regression-aware prompt optimisation for LLM judges. 
Adapting dynamic prompt-tuning frameworks~\cite{khattab2024dspy,yang2024large} could automatically adjust constraints across model versions to explicitly minimise capability loss.